\documentclass[prl,twocolumn,showpacs,preprintnumbers,amsmath,epsf]{revtex4}
\usepackage{graphicx}
\usepackage{dcolumn}
\usepackage{bm}

\begin{document}

\title{LIQUID-LIKE PHASES OF $\bm{\pi^+\pi^-}$ MATTER}

\author{D.V. Anchishkin}\email{anch@bitp.kiev.ua}
\author{A.V. Nazarenko}\email{nazarenko@bitp.kiev.ua}
\affiliation{Bogolyubov Institute for Theoretical Physics,\\
14-b, Metrologichna Str., Kiev 03143, Ukraine}

\date{\today}

\begin{abstract}
To give a common theoretical description of liquid phases of the
charged pion matter in a wide temperature interval, the relativistic
quantum $\varphi^6$ type model is considered. The liquid states
of pion condensate and hot pion matter are investigated.
\end{abstract}



\maketitle

\section{INTRODUCTION}

The possibility to observe the pion condensate in heavy-ion collisions
had served as a subject of theoretical and experimental investigations
during last decades. This condensed state is often imaged like multipion
droplets or/and pion liquid. The pion condensate is also a necessary
component of neutron stars. The theoretical description of pion condensate
is usually given within the framework of phenomenological models leading to
the $\varphi^4$ type self-interaction between pionic degrees of freedom
(see, for example, \cite{MVST,MG,SH1}).

Since the pions are the Goldstone bosons for spontaneously broken chiral
symmetry, a realistic description of pion subsystem can only be
achieved on the basis of a model respecting chiral symmetry. However, if the
pion matter is at the temperature much lower than the temperature of the
chiral phase transion ($T_\chi\approx 150$~MeV), a chiral perturbation theory
is applicable. Using perturbation scheme, we would like to point out the general
property of the chiral models, namely, in third-order approximation they result
in an attractive two-body interaction (associated with $\varphi^4$ term) and
three-body repulsive interaction (associated with $\varphi^6$ term). To
demonstrate it, we appeal to the Skyrme~\cite{Sk} and Weinberg~\cite{Wein}
models, where interactions can be presented as
$V_{\rm Skyrme}(\nu)\equiv-f^2_\pi m^2{\rm Tr}(U+U^\dag-2)/4
=f^2_\pi m^2(\nu^2/2!-\nu^4/4!+\nu^6/6!-\ldots)$ and $V_{\rm Weinberg}(\nu)\equiv
f^2_\pi m^2 \nu^2/(2+\nu^2/2)=f^2_\pi m^2(\nu^2/2-\nu^4/8+\nu^6/32-\ldots)$,
respectively. Here we use notations for
$U=\exp{(i\vec\tau\cdot\vec\pi/f_\pi)}\in SU(2)$,
$\nu^2={\vec\pi}^2/f^2_\pi$; $f_\pi$ and $m$ are pion decay constant and
pion mass, respectively. Note that these interactions are
the limited functions of $\nu$. Also note that the theoretical
Gasser-Leutwyler interaction~\cite{GL} of order $p^4$ is attractive,
and the experimental data justify the existence of three-pion correlations in
heavy-ion collisions~\cite{star}. Therefore, if we limit ourselves by
consideration of the third-order approximation (leading to the the
relativistic quantum $\varphi^6$ type model), we can expect to observe
first-order phase transitions (PTs) in pion subsystem.

Indeed, it is already known from molecular physics that the non-relativistic
$\varphi^6$ model, in contrast to the $\varphi^4$ one, allows us to observe
not only second-order PT into condensate state but also its
gaseous and liquid phases. It is reached by means of solving the
corresponding Gross-Pitaevskii equation \cite{GFTC,KT}. Remark that the
$\varphi^6$ model has been already used in nuclear hydrodynamics and quantum
field theory~\cite{BM}. Here we are trying to obtain similar results
(without degrees of freedom of $\pi^0$ mesons) in the context of the physics
of superdense ions and neutron stars, where pion condensate plays an
important role in the softening of nucleonic equation of state~\cite{MVST}.
Clearly, in this problem, we should take into account the essentially
different nature of interactions in molecular physics and pion subsystem.
Moreover, we would like to investigate the conditions of existence of
liquid phase at high temperatures (at the temperatures higher than the
temperature $T_{\rm cond}$ of the second-order PT into condensate state) with the
use of the same model. It is possible that the hot $\pi^+\pi^-$ liquid can
be created in relativistic heavy-ion collisions, where the pions play a
dominant role at the final stage of the reaction. Note that collective
phenomena in particle-nucleus and nucleus-nucleus collisions are
well-established and play significant role. An existence of
the hot pion liquid has been already predicted in Ref.~\cite{KGSG}.

Since the liquid condensate of $\pi^+\pi^-$ mesons thermodynamically differs
from gaseous condensate, then such a difference should be taken into
account, when attempts to register an appearance of a condensate in
heavy-ion collisions are performed. On the other hand, the system of the
large number of pions at high temperatures, as we shall see below, can also
be in a liquid phase which can essentially affect on nucleonic dynamics. We
say ``liquid'', when we deal with a dense phase in thermodynamics. However,
the same ``liquid'' is a state with a high magnitude of pionic field from
the field-theoretical point of view. As a consequence, amplification of
pionic field can result in the creation of a proton-antiproton pair.

\section{THE PHASES OF PION CONDENSATE}

 The Lagrangian density ${\cal L}(\pi^\dag,\pi)$ of the model is
\begin{equation}\label{eq1}
{\cal L}=\partial_\mu\pi^\dag\partial^\mu\pi
-m^2\pi^\dag\pi
+\frac{A}{2!}(\pi^\dag\pi)^2
-\frac{B}{3!}(\pi^\dag\pi)^3,
\end{equation}
 where the normal ordering of operator fields is assumed. We deal with the
 case, when electromagnetic interaction is neglected. Here it is also
 supposed that $A=m^2/g^2$, $B=3m^2\lambda/2g^4$, $m=140$~MeV, $g$ and
 $\lambda$ are model parameters which should be fitted.

 Constant $g$ plays a role of pion decay constant $f_\pi$ redefined in
 medium. Model parameter $\lambda$ is introduced to account effectively the
 higher-order terms of chiral interaction expansion. To analyze the range of
 $\lambda$, let us present chiral interactions (considered above, for
 example) as $V(\nu)=a_2\nu^2-a_4\nu^4+a_6\nu^6\lambda(\nu)$, where
 $a_n\equiv |V^{(n)}(0)|/n!$ and $\nu^2=2\pi^\dag\pi/g^2$. It turns
 out that $\lambda(\nu)$ is a smooth function, which is decreased from 1 to
 0, when $\nu$ runs from 0 to $\infty$. Thus, if even $|\nu|>1$, the
 constant $\lambda$ introduced instead of function $\lambda(\nu)$ should
 be less than 1 in order to relate our phenomenological approach with
 the chiral theory.

 Our approach to this model is based on {\it ansatz} on applicability
 of the mean-field approximation (MFA) in a wide temperature interval.
 Moreover, we are limiting ourselves by two cases: i) $T=0$, when the
 operator field $\pi$ can be replaced by a classical complex field $\phi$,
 also called as the order parameter, describing a condensate; ii)
 $T>T_{\rm cond}$, when there are no anomalous expectation values and the field
 $\pi$ coincides with quantum fluctuations $\chi$. Our aim is to investigate
 the liquid-like states of $\pi^+\pi^-$ matter in these two regimes.

 It is appropriate for us to begin from the investigation of different phases
 of the non-uniform pion condensate at $T=0$, which appears in the neutron
 stars and the nuclei with density higher than saturation one. This
 situation is modeled by Lagrangian density ${\cal L}(\phi^*,\phi)$. In
 fact the replacement of $\pi$ by $\phi$ is analogous to the transition
 from quantum electrodynamics to the classical description of
 electromagnetism, when a big number of photons are in approximately the
 same state. In our case, the presence of a big number of pions in a single
 state (Bose-Einstein condensate) permits us to introduce the classical
 function $\phi$. In a contrast to Maxwell theory, ${\cal L}(\phi^*,\phi)$
 contains the quantum constant $\hbar$ explicitly ($\hbar=1$ in our units).
 On the other hand, the quantum meaning of classical fields is reviewed in
 Ref.~\cite{Jackiw}.

 From the variational derivative of the corresponding classical action
 functional, one obtains the following evolution equation:
\begin{equation}\label{eq3}
\left(\frac{\partial^2}{\partial t^2}-\bm{\nabla}^2\right)\phi
+m^2\phi-A|\phi|^2\phi+\frac{B}{2}|\phi|^4\phi=0,
\end{equation}
 which serves as a relativistic generalization of the corresponding
 Gross-Pitaevskii equation without external trap potential~\cite{GFTC}.

 An applicability of Eq.~(\ref{eq3}) demands that the following conditions
 should be satisfied. First, the total number of pions (multiplicity) must
 be large enough because only in this case we are authorized to use the
 concept of Bose-Einstein condensation. Second, in order to replace the
 field operator by the classical field we have to assume both
 diluteness and the fact that the temperature is low enough. This allows us to
 ignore both the quantum and thermal depletion of the condensate.

 The solution of Eq.~(\ref{eq3}) essentially depends on physical
 parameters and boundary conditions. Here we shall deal with exact static
 quasi one-dimensional solution (``bubble'') which can be presented as:
\begin{equation}\label{eq4}
\phi(t,z)=\eta_k(z)\exp{\left(-imt\sqrt{1-k^2}\right)},
\end{equation}
 where $\eta_k(z)$ is a real function.

 The time dependence is in a phase, corresponding to the non-vanishing charge
 $Q$. Such a substitution results in a very large class of solutions
 $\eta_k(z)$ in terms of elementary and elliptic functions~\cite{SS,GTW}.
 However, we are interested in stable solutions arising from ordinary
 charge conservation.

 For the boundary conditions $\lim\limits_{z\to\pm\infty}\eta_k(z)=0$,
 $\lim\limits_{z\to\pm\infty}\eta^{\prime}_k(z)=0$, there is a solitary
 wave solution of the form~\cite{SS}:
\begin{equation}
\eta_k(z)=\frac{2gk}{\sqrt{\sqrt{1-4\lambda k^2}
{\rm cosh}(2kmz)+1}}.
\end{equation}
 The result obtained describes a narrow layer of the dense phase trapped
 between dilute phases and leads to a localized energy density. This model
 solution can describe a region of dense pion matter with respect to the axis
 of nucleus-nucleus collision.

 One can simply prove that the solution (\ref{eq4}) is pseudoscalar:
 it is necessary to replace the coordinate $z$ and the (quasi)momentum $k$
 by $-z$ and $-k$, respectively. As it must be, this operation changes
 a sign of $\phi(t,z)$.

 An order parameter in this system is the scalar density
 $n(z)\equiv|\phi(t,z)|^2$. To determine the phase of pion matter in dense
 region, we introduce the charge $Q$ (difference between the numbers of
 particles and antiparticles) and the total energy $E$ per unit area:
\begin{eqnarray}
Q&=&8g^2\frac{\sqrt{1-k^2}}{\sqrt{\lambda}}
{\rm arctanh}{\sqrt{\frac{1-\sqrt{1-4\lambda k^2}}{1+\sqrt{1-4\lambda k^2}}}},
\\
E&=&m\sqrt{1-k^2}Q+\frac{g^2mk}{\lambda}
\nonumber\\
& &-\frac{g^2m}{\lambda^{3/2}}
(1-4\lambda k^2)
{\rm arctanh}\frac{1-\sqrt{1-4\lambda k^2}}{2k\sqrt{\lambda}}.
\end{eqnarray}
 These quantities are parameterized by $k$ as independent variable. Also,
 it is not hard to see that, varying $k$, the behavior of functions $Q$,
 $E$ is mainly determined by parameter $\lambda$. At this time, the model
 parameter $g$ influences only the magnitude of these characteristics.

 Since the charge $Q$ is conserved in time, there exists a non-vanishing
 chemical potential $\mu\equiv\partial E/\partial Q$, which is calculated by
 the formula: $\mu=(\partial E/\partial k)/(\partial Q/\partial k)$. The
 dependence of $\mu$ on $k$ can be explicitly found. However, we do not
 adduce it here because of its cumbersome form. Note only that $\mu>0$. It
 is in accordance with the result of \cite{MG}, where multipion
 droplets are considered at low temperature.

 As shown in Fig.~1, the central density $\nu\equiv n(0)/g^2$ and
 the chemical potential $\mu$ present backbendings typical for the first-order
 PT. The transition point, given by the crossing point in
 $\varepsilon\equiv E/mg^2$ versus $q\equiv Q/g^2$, corresponds to a Maxwell
 construction in the diagram of $\mu$ versus $q$. However, the system should
 never explore the backbending part of the diagram because it is a
 metastable state. It is clear that dense phase is associated with a liquid
 while dilute phase is a gas. We want to stress that both branches are
 quantum fluids.

 We see in Fig.~1 that the first-order PT in pion condensate takes place at
 $0.25<\lambda<0.268$ and $\lambda<1$ as is argued above. Note that, at
 $\lambda\approx0.268$, the stable and metastable solutions coincide. It
 defines a critical point associated with a second-order PT: at this point
 the derivative of $\nu$ as a function of $q$ diverges. For $\lambda=0.25$
 the attractive two-body interaction prevails and the system tends to
 collapse. In this case the maximum charge is limited by $q\approx 5.3$.
 Assuming an existence of the first-order phase transition, we put
 $\lambda=0.26$ in this paper. The value of constant $g$ will be discussed
 in the case of hot pion matter.

\begin{figure}[htbp]
\begin{picture}(60,45)
\put(-90.4,61){\includegraphics[width=6cm,angle=-90]{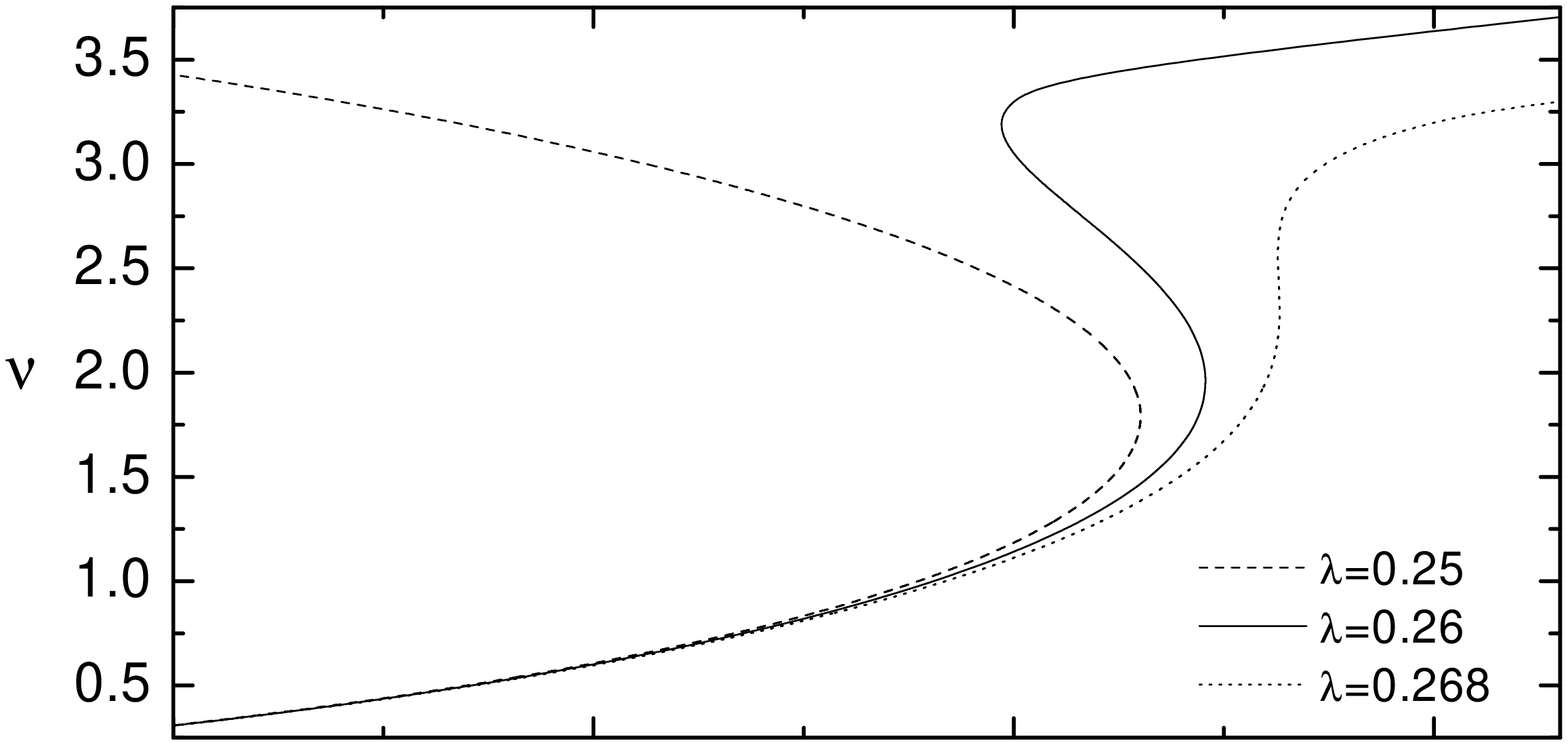}}
\put(-90,-35){\includegraphics[width=6cm,angle=-90]{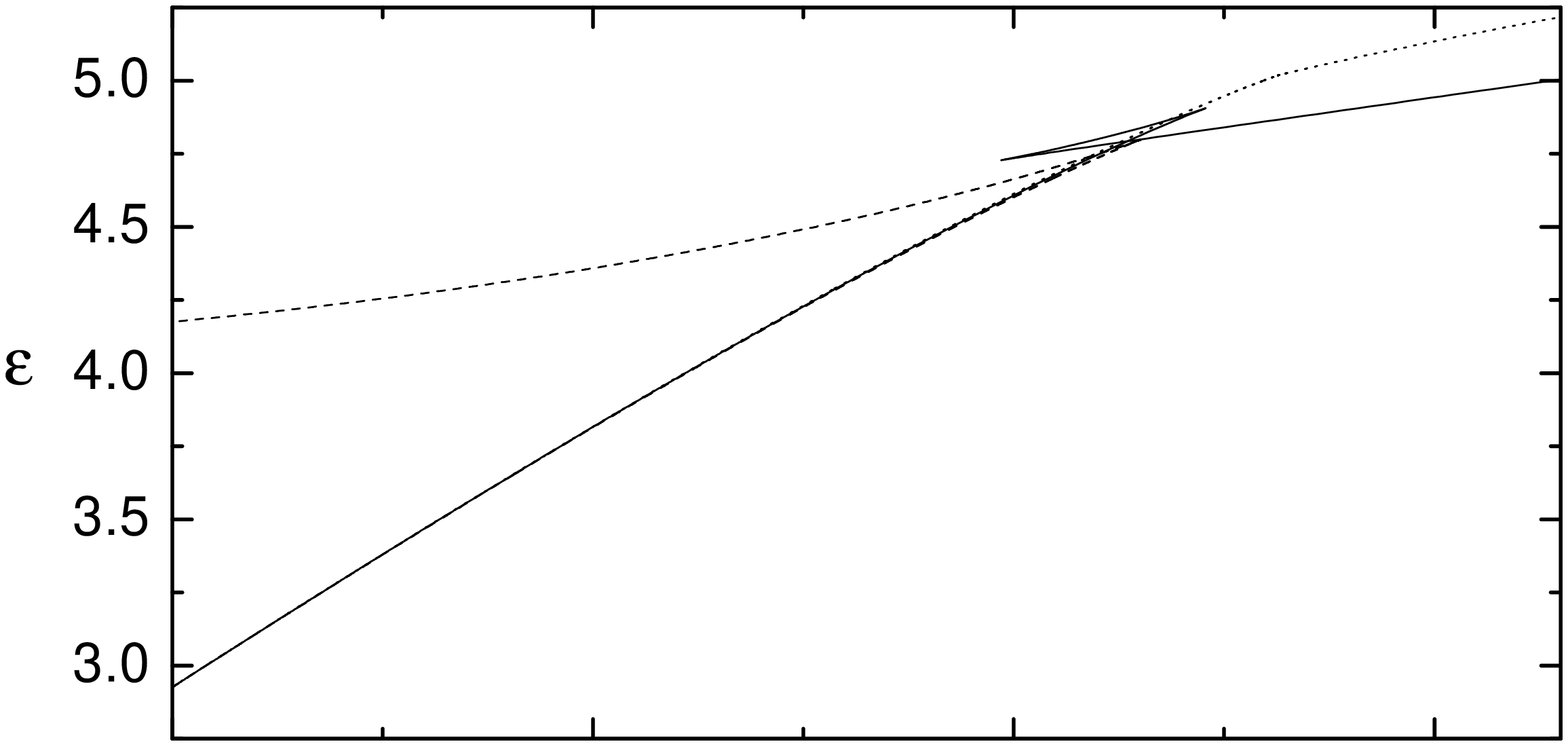}}
\put(-90,-131){\includegraphics[width=6cm,angle=-90]{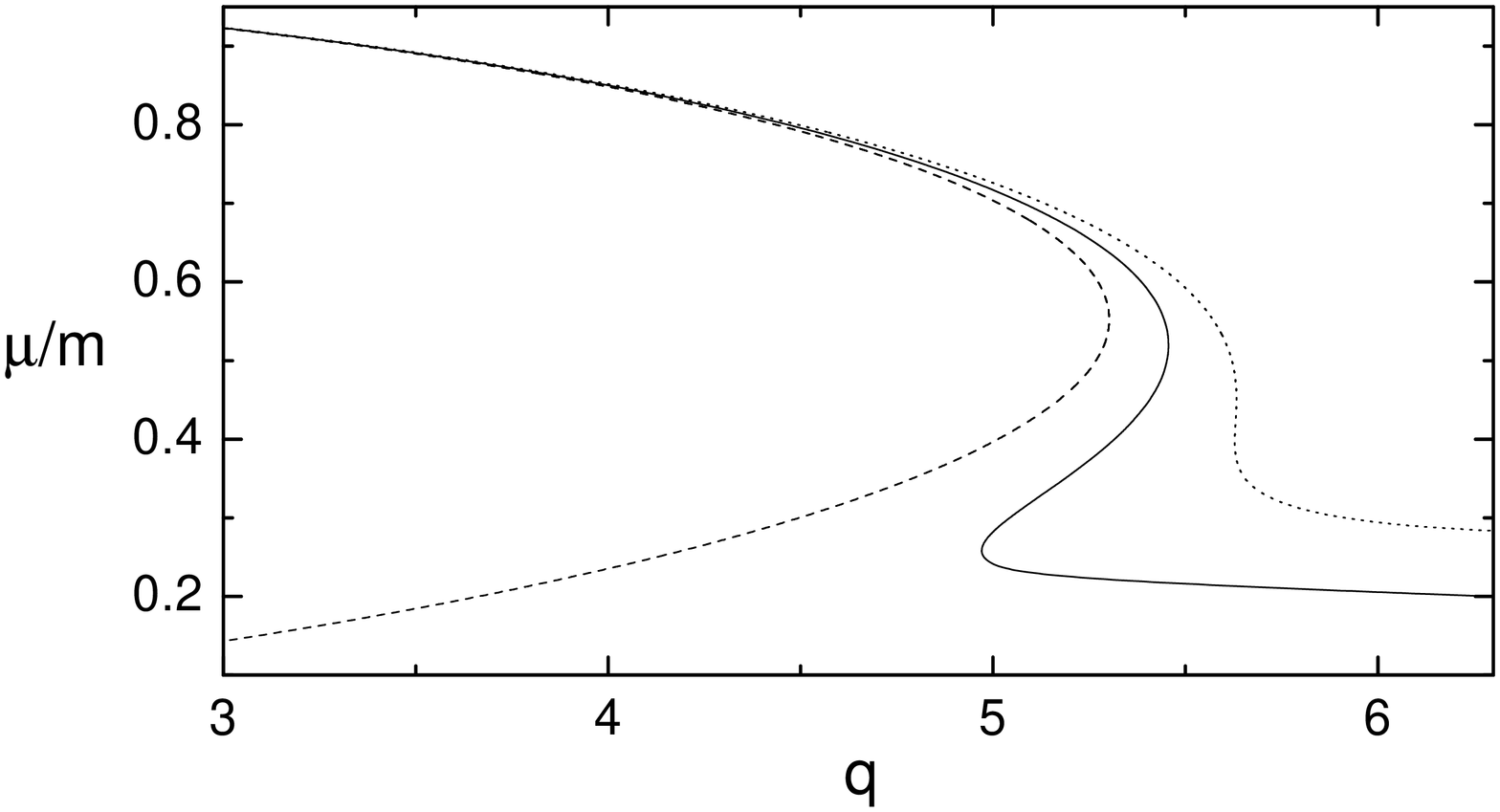}}
\end{picture}
\vspace*{90mm}
\caption{\label{fig1} Central density $\nu$, total energy $\varepsilon$,
chemical potential $\mu$ per mass $m$, in dimensionless units, as functions
of the reduced charge $q$.}
\end{figure}

\section{THE HOT PION LIQUID}

 Now we focus on the description of the $\pi^+\pi^-$ matter at high
 temperatures in MFA, when the field $\pi$ is described completely by
 quantum fluctuations $\chi$. We represent operator field $\chi$ in finite
 volume $V$ as
\begin{eqnarray}
\chi(x)&=&\sum\limits_{\bf p}\left[a_{\bf p}u_{\bf p}(x)
+b^\dag_{\bf p}u^*_{\bf p}(x)\right],
\label{sol1}\\
u_{\bf p}(x)&=&\frac{1}{\sqrt{2E_{\bf p}V}} \exp{(-iE_{\bf
p}t+i{\bf px})}.
\end{eqnarray}
 The dispersion law is $E_{\bf p}=\sqrt{{\bf p}^2+m^2_*}$, where $m_*$ is
 effective mass which should be found; $a^\dag_{\bf p}$, $b^\dag_{\bf p}$
 ($a_{\bf p}$, $b_{\bf p}$) are creation (annihilation) operators of
 particles and antiparticles, respectively. The commutation relations
 between operators are standard:
 $[a_{\bf p},a^\dag_{\bf q}]=\delta_{{\bf p},{\bf q}}$,
 $[b_{\bf p},b^\dag_{\bf q}]=\delta_{{\bf p},{\bf q}}$.

 The Hamiltonian operator, $H=\int({\cal H}_0+{\cal W})d^3x$, based on
 ${\cal L}(\chi^\dag,\chi)$, is determined by the following terms:
\begin{eqnarray}
{\cal H}_0&=&\partial_t\chi^\dag\partial_t\chi+
\bm{\nabla}\chi^\dag\bm{\nabla}\chi+m^2_*\chi^\dag\chi,\\
{\cal W}&=&(m^2-m^2_*)\chi^\dag\chi
-\frac{A}{2!}(\chi^\dag\chi)^2
+\frac{B}{3!}(\chi^\dag\chi)^3.
\end{eqnarray}
 One has
\begin{equation}
H=\sum\limits_{\bf p}E_{\bf p}(n_{\bf p}+{\bar n}_{\bf p})+\int{\cal W}d^3x,
\end{equation}
 where $n_{\bf p}\equiv a^\dag_{\bf p}a_{\bf p}$ and
 ${\bar n}_{\bf p}\equiv b^\dag_{\bf p}b_{\bf p}$ are the number
 operators of particles and antiparticles, respectively.

 Assuming that the interaction term is the perturbation, thermodynamic potential can be
 written as
\begin{equation}
\Omega=T\sum\limits_{\gamma,{\bf p}}\ln
\left[1-{\rm e}^{\beta (\gamma\mu-E_{\bf p})}\right]
+V{\cal W}_{\rm MFA},
\end{equation}
 where $\mu$ is the chemical potential, $\beta=1/T$ is inverse temperature,
 $\gamma=\pm1$.

 In our approximation, the interaction term is as follows
\begin{equation}
{\cal W}_{MFA}=(m^2-m^2_*)\sigma+m^2(b\sigma^3-2a\sigma^2),
\end{equation}
 where $\sigma=\langle\chi^\dag\chi\rangle$ is the order parameter,
 $a=1/2g^2$, $b=3\lambda/2g^4$.

 We find thermodynamically conjugate variables $\sigma$ and $m_*$ from the
 extremizing conditions $\partial\Omega/\partial m_*=0$,
 $\partial\Omega/\partial\sigma=0$, which result in the gap equation,
\begin{equation}
\sigma=\frac{1}{V}
\sum\limits_{\gamma,{\bf p}}
\frac{1}{2E_{\bf p}}
\frac{1}{{\rm e}^{\beta (E_{\bf p}-\gamma\mu)}-1},
\end{equation}
 and the mass ratio, $s^2\equiv m^2_*/m^2=1-4a\sigma+3b\sigma^2$.

 Further investigation of thermodynamic properties of hot pion matter
 is carried out in the thermodynamic limit, when the replacement of summation
 by integration over momentum takes place. Here we neglect a contribution
 of the surface.

 It turns out that, to solve correctly the constraint, it is helpful to
 operate with dimensionless variable $s$ instead of $\sigma$. It leads at
 once to the splitting of the value interval of $\sigma$ into two branches:
 $\sigma=(2g^2/9\lambda)[1\pm\sqrt{1-(9/2)\lambda(1-s^2)}]$. Such a
 representation of $\sigma$ gives us two equations $F_\pm(s,T,\mu)=0$,
 where
\begin{eqnarray}
&&F_\pm(s,T,\mu)=1\pm\sqrt{1-\frac{9}{2}\lambda(1-s^2)}
-\frac{\lambda}{2}\left(\frac{3m}{2\pi g}\right)^2s^2
\nonumber\\
&&\times
\sum\limits_{\gamma}
\int\limits_0^\infty \frac{dk}{\sqrt{1+k^2}}
\frac{k^2}{{\rm e}^{\beta\left(ms\sqrt{1+k^2}-\gamma\mu\right)}-1}.
\end{eqnarray}
 The solution of these equations is the function $s(T,\mu)$ which is
 calculated numerically.

 One can say that this splitting reflects the existence of two phases of
 hot pion matter. If $s=1$, we find that a single phase, corresponding
 to ideal gas, survives. Switching on an interaction ($s\not=1$), a new
 phase, called as hot pion liquid, can arise. The appearance of this phase
 is possible due to the sixth-order term.

\begin{figure}[htbp]
\begin{picture}(60,45)
\put(-90, 52){\includegraphics[width=6cm,angle=-90]{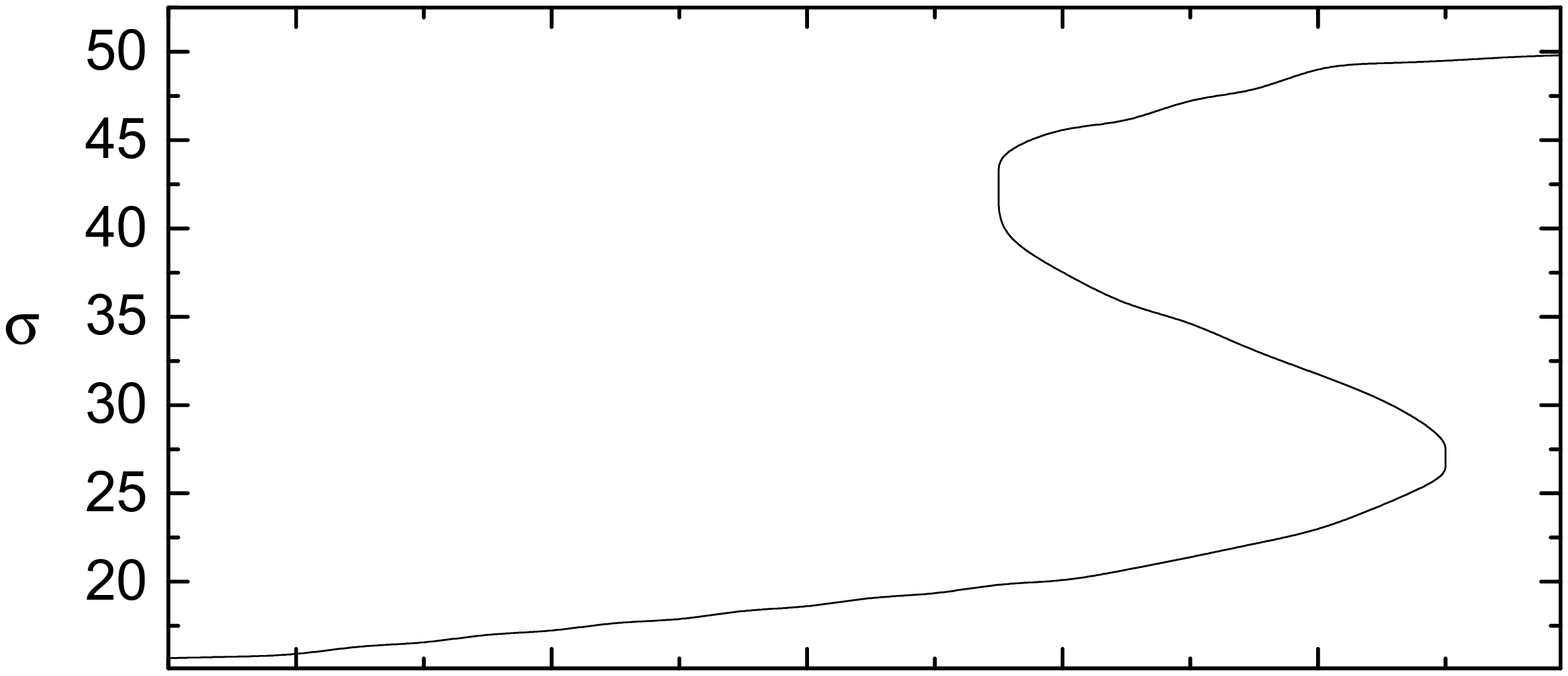}}
\put(-90,-35){\includegraphics[width=6cm,angle=-90]{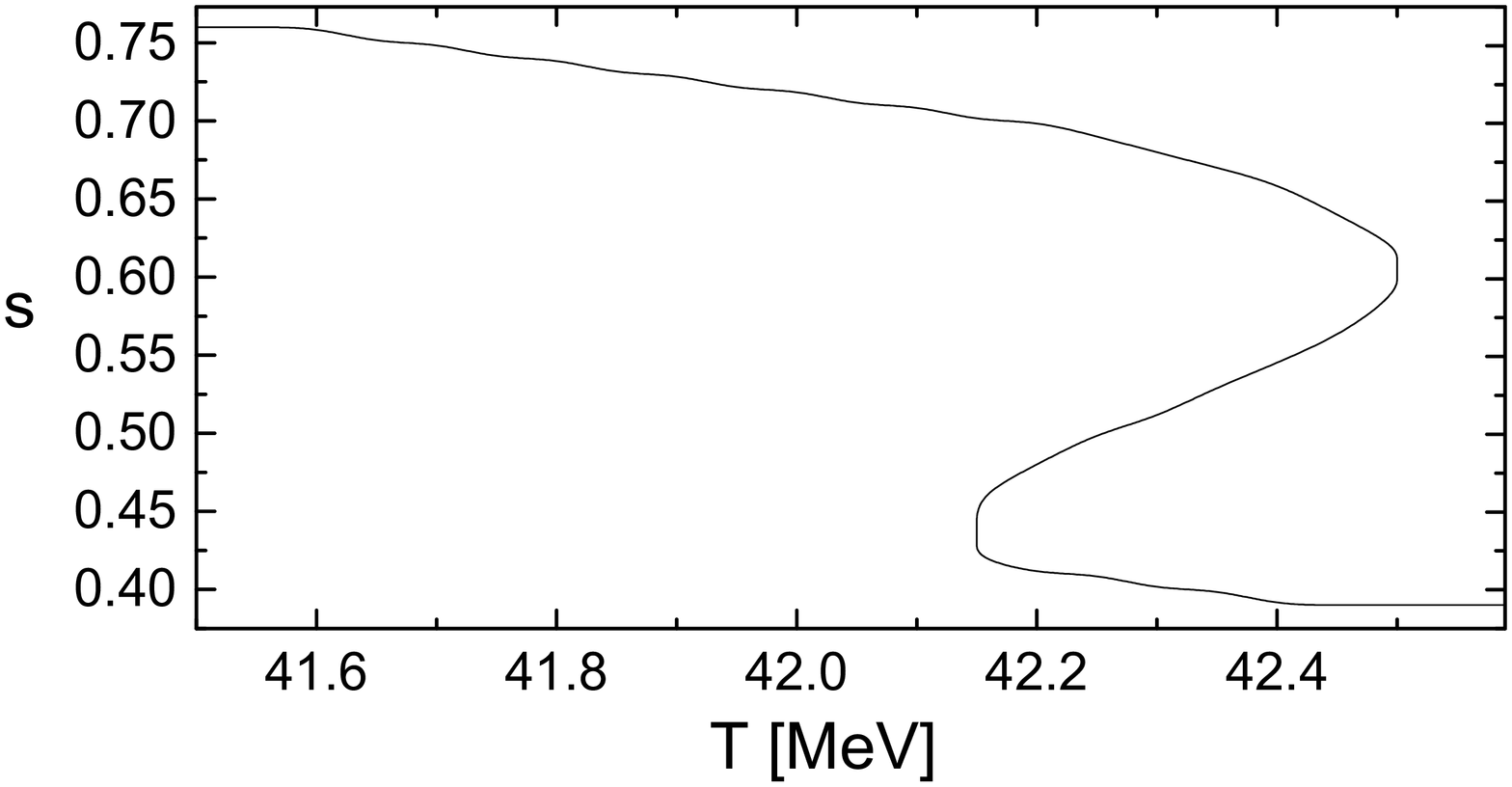}}
\end{picture}
\vspace*{50mm}
\caption{\label{fig2} The dependence of the scalar density
$\sigma$ (in MeV${}^2$) and the reduced effective mass $s\equiv m_*/m$ on
temperature.}
\end{figure}

 In order to compute thermodynamic functions, one needs to know the values
 of two parameters of our model, namely, $g$ and $\lambda$. We have
 already assumed that $\lambda=0.26$ to achieve liquid-gas phase transition
 in condensate. To fit constant $g$, we try to appeal to experimental data
 obtained by DLS Collaboration in Berkeley and extract necessary
 information from dilepton spectra produced in proton-nucleus
 reaction~\cite{DLS}. If we believe that the main contribution to the
 spectrum comes from hot $\pi^+\pi^-$ annihilation, one can conclude that
 the pion annihilation threshold is not equal to twice free pion mass
 ($2m$) but smaller and approximately is $2ms=260$~MeV. Our fit to the
 spectra~\cite{DLS}, which depends on the dielectron invariant mass, gives
 us the temperature of the fireball, formed in this collision, equal to
 $T=75$~MeV. Therefore, to find $g$ at chemical equilibrium ($\mu=0$), we
 need to substitute these data in our equations. One obtains that
 $g\approx8$~MeV from equation $F_+(s,T,0)=0$. Using the equation
 $F_-(s,T,0)=0$, another value of $g$ is derived. However, the liquid-gas
 PT is not observed in this case.

 Assuming chemical equilibrium, there is the only possibility
 to observe the transition into liquid phase with changing temperature.
 Indeed, the dense ``liquid'' phase appears with increasing temperature (see
 Fig.~2) that is in contrast to molecular physics, where the number of
 particles is conserved. This phenomenon has been already pointed out in
 \cite{KGSG}. The critical temperature of the PT in our model is
 about $42.3$~MeV and less than $136$~MeV as it was predicted in
 \cite{KGSG}. However, we should conclude that, at $T=75$~MeV of fireball,
 the pions are in the liquid phase, if the scenario, when $g\approx8$~MeV, is
 realized in nature. This outcome demands additional theoretical and
 experimental verifications.

 Note that the calculations with non-vanishing chemical potential can be
 carried out and will be published elsewhere.

\section*{ACKNOWLEDGMENTS}

A.N. is greatly indebted to D.N.~Voskresensky (GSI, Darmstadt) for fruitful
discussions and V.B.~Blavatska (ICMP, Lviv) for obtaining phase diagrams.


\end{document}